\documentclass[12pt]{article}
\usepackage{sw20jart}
\usepackage{amsmath}

\newtheorem{theorem}{Theorem}[section]

\newtheorem{lemma}[theorem]{Lemma}

\newtheorem{remark}[theorem]{Remark}

\newenvironment{proof}[1][Proof]{\noindent\textbf{#1.} }{\ \rule{0.5em}{0.5em}}
\input{tcilatex}

\begin{document}

\author{Mark Korenblit \\
Holon Institute of Technology, Israel\\
korenblit@hit.ac.il\bigskip \\
Vadim E. Levit\\
Ariel University, Israel\\
levitv@ariel.ac.il }
\title{Decomposition Lemmas}
\date{}
\maketitle

\section{The First Decomposition Lemma}

We define the following recursive function $T(n)$: 
\begin{eqnarray}
T(0) &=&0  \notag \\
T(1) &=&0  \notag \\
T(2) &=&1  \notag \\
T(n) &=&T\left( \left\lceil \frac{n}{2}\right\rceil \right) +T\left(
\left\lfloor \frac{n}{2}\right\rfloor +1\right) +T\left( \left\lceil \frac{n%
}{2}\right\rceil -1\right) +T\left( \left\lfloor \frac{n}{2}\right\rfloor
\right) +1\text{\qquad \qquad }  \label{f2} \\
(n &\geq &3).  \notag
\end{eqnarray}

\begin{lemma}
\label{lem_n/2_a}For $n\geq 1$, $1\leq i\leq n$%
\begin{equation}
T(n)\leq T(i)+T(n-i+1)+T(i-1)+T(n-i)+1.  \label{f1}
\end{equation}%
In addition, for $n\geq 3$ 
\begin{equation}
T(n)=T(i)+T(n-i+1)+T(i-1)+T(n-i)+1  \label{f13}
\end{equation}%
if and only if $i$ is equal to $\frac{n+1}{2}$ for odd $n$ and $i$ is equal
to $\frac{n}{2}$ or $\frac{n}{2}+1$ for even $n$.
\end{lemma}

\proof%
Substitution of $\frac{n+1}{2}$ (odd $n$) and $\frac{n}{2}$ or $\frac{n}{2}%
+1 $ (even $n$) for $i$ in (\ref{f13}) reduces the right part of (\ref{f13})
to the right part of (\ref{f2}), i.e., the direct assertion for relation (%
\ref{f13}) holds by the definition of $T(n)$.

Now, separately, consider special cases when $i=1$ and when $i=n$. In both
cases 
\begin{eqnarray}
&&T(i)+T(n-i+1)+T(i-1)+T(n-i)+1  \notag \\
&=&T(1)+T(n)+T(0)+T(n-1)+1>T(n)  \label{f14}
\end{eqnarray}%
and, therefore, (\ref{f1}) holds. Specifically, for $n=1$ and $n=2$ the
single possible values of $i$ are $1$ and $n$. Hence, the lemma is proven
for these values of $n$.

The proof of (\ref{f1}) for $n\geq 3$, $2\leq i\leq n-1$ and of the opposite
assertion for (\ref{f13}) is based on mathematical induction. It can be
shown that (\ref{f1}) holds for $n=3$ and $n=4$. For $n=3$ the single
possible value of $i$ is $2$, i.e., $\frac{n+1}{2}$ and for $n=4$ the
possible values of $i$ are $2$ and $3$, i.e., $\frac{n}{2}$ and $\frac{n}{2}%
+1$, respectively. Since (\ref{f13}) holds for these values of $i$ then (\ref%
{f1}) holds for them also. Besides, as follows from (\ref{f14}), (\ref{f13})
is correct only for these values of $i$. By the way, based on (\ref{f2}),
the values of $T(3)$ and $T(4)$ are $3$ and $6$, respectively. Now we will
prove the lemma for any $n>4$ on condition that it is correct for $%
1,2,3,4,5,\ldots ,n-1$.

As follows from (\ref{f2}), in the case when $i-1$ and $n-i$ are not less
than $3$ the following equations hold: 
\begin{eqnarray}
T(i) &=&T\left( \left\lceil \frac{i}{2}\right\rceil \right) +T\left(
\left\lfloor \frac{i}{2}\right\rfloor +1\right) +  \notag \\
&&T\left( \left\lceil \frac{i}{2}\right\rceil -1\right) +T\left(
\left\lfloor \frac{i}{2}\right\rfloor \right) +1  \label{f19} \\
T(n-i+1) &=&T\left( \left\lceil \frac{n-i+1}{2}\right\rceil \right) +T\left(
\left\lfloor \frac{n-i+1}{2}\right\rfloor +1\right) +  \notag \\
&&T\left( \left\lceil \frac{n-i+1}{2}\right\rceil -1\right) +T\left(
\left\lfloor \frac{n-i+1}{2}\right\rfloor \right) +1\qquad \qquad
\label{f20} \\
T(i-1) &=&T\left( \left\lceil \frac{i-1}{2}\right\rceil \right) +T\left(
\left\lfloor \frac{i-1}{2}\right\rfloor +1\right) +  \notag \\
&&T\left( \left\lceil \frac{i-1}{2}\right\rceil -1\right) +T\left(
\left\lfloor \frac{i-1}{2}\right\rfloor \right) +1  \label{f21} \\
T(n-i) &=&T\left( \left\lceil \frac{n-i}{2}\right\rceil \right) +T\left(
\left\lfloor \frac{n-i}{2}\right\rfloor +1\right) +  \notag \\
&&T\left( \left\lceil \frac{n-i}{2}\right\rceil -1\right) +T\left(
\left\lfloor \frac{n-i}{2}\right\rfloor \right) +1  \label{f22}
\end{eqnarray}%
Suppose, for the moment, that these equations hold always. In such a case,
the proof of (\ref{f1}) could be organized as follows. We collect the
components of the right parts (except the units) of equations (\ref{f19} -- %
\ref{f22}) to the following square matrix of size four: 
\begin{equation}
\left( 
\begin{array}{llll}
T\left( \left\lceil \frac{i}{2}\right\rceil \right) & T\left( \left\lfloor 
\frac{i}{2}\right\rfloor +1\right) & T\left( \left\lceil \frac{i}{2}%
\right\rceil -1\right) & T\left( \left\lfloor \frac{i}{2}\right\rfloor
\right) \\ 
T\left( \left\lceil \frac{n-i+1}{2}\right\rceil \right) & T\left(
\left\lfloor \frac{n-i+1}{2}\right\rfloor +1\right) & T\left( \left\lceil 
\frac{n-i+1}{2}\right\rceil -1\right) & T\left( \left\lfloor \frac{n-i+1}{2}%
\right\rfloor \right) \\ 
T\left( \left\lceil \frac{i-1}{2}\right\rceil \right) & T\left( \left\lfloor 
\frac{i-1}{2}\right\rfloor +1\right) & T\left( \left\lceil \frac{i-1}{2}%
\right\rceil -1\right) & T\left( \left\lfloor \frac{i-1}{2}\right\rfloor
\right) \\ 
T\left( \left\lceil \frac{n-i}{2}\right\rceil \right) & T\left( \left\lfloor 
\frac{n-i}{2}\right\rfloor +1\right) & T\left( \left\lceil \frac{n-i}{2}%
\right\rceil -1\right) & T\left( \left\lfloor \frac{n-i}{2}\right\rfloor
\right)%
\end{array}%
\right)  \label{f4}
\end{equation}%
We want to show that $T(n)$ does not exceed the sum of all elements of this
matrix plus five (four units from equations (\ref{f19} -- \ref{f22}) and in
addition, the unit from (\ref{f1})). To that end, we reveal four groups of
elements in the matrix so that $T\left( \left\lceil \frac{n}{2}\right\rceil
\right) $, $T\left( \left\lfloor \frac{n}{2}\right\rfloor +1\right) $, $%
T\left( \left\lceil \frac{n}{2}\right\rceil -1\right) $, and $T\left(
\left\lfloor \frac{n}{2}\right\rfloor \right) $, respectively, do not exceed
the sum of elements in the corresponding group plus $1$. We do it in the
following way. We denote the feasible partition of the number $2n$ to four
numbers $i$, $n-i+1$, $i-1$, and $n-i$ ($1<i<n$) by $\varepsilon $\textit{%
-partition}. Note that the arguments in the right part of (\ref{f1}) obey
the $\varepsilon $-partition. And now, we divide the whole set of the matrix
elements into four certain groups each of size four (\textit{quartets}).
These quartets are chosen in such a way that the arguments of each of the
elements in the four quartets represent the $\varepsilon $-partition of $%
2\left\lceil \frac{n}{2}\right\rceil $, $2\left( \left\lfloor \frac{n}{2}%
\right\rfloor +1\right) $, $2\left( \left\lceil \frac{n}{2}\right\rceil
-1\right) $, and $2\left\lfloor \frac{n}{2}\right\rfloor $, respectively.
Note that since $\left\lceil \frac{n}{2}\right\rceil $, $\left\lfloor \frac{n%
}{2}\right\rfloor +1$, $\left\lceil \frac{n}{2}\right\rceil -1$, and $%
\left\lfloor \frac{n}{2}\right\rfloor $ are less than $n$, such a grouping
gives the desired four groups. The sums of the elements of each of these
groups added by $1$ are not less than $T\left( \left\lceil \frac{n}{2}%
\right\rceil \right) $, $T\left( \left\lfloor \frac{n}{2}\right\rfloor
+1\right) $, $T\left( \left\lceil \frac{n}{2}\right\rceil -1\right) $, and $%
T\left( \left\lfloor \frac{n}{2}\right\rfloor \right) $, respectively.

The possibility of the required fragmentation of the matrix (\ref{f4}) has
to hold for both even and odd $n$ and for both even and odd $i$. Hence, four
transformations of the initial matrix (\ref{f4}) should be considered. These
transformations can be written in the following way:

Even $n$, even $i$: 
\begin{equation}
\left( 
\begin{array}{cccc}
\frac{i}{2} & \frac{i}{2}+1 & \frac{i}{2}-1 & \frac{i}{2} \\ 
\frac{n-i}{2}+1 & \frac{n-i}{2}+1 & \frac{n-i}{2} & \frac{n-i}{2} \\ 
\frac{i}{2} & \frac{i}{2} & \frac{i}{2}-1 & \frac{i}{2}-1 \\ 
\frac{n-i}{2} & \frac{n-i}{2}+1 & \frac{n-i}{2}-1 & \frac{n-i}{2}%
\end{array}%
\right)  \label{f5}
\end{equation}

Odd $n$, odd $i$: 
\begin{equation}
\left( 
\begin{array}{cccc}
\frac{i+1}{2} & \frac{i+1}{2} & \frac{i-1}{2} & \frac{i-1}{2} \\ 
\frac{n-i}{2}+1 & \frac{n-i}{2}+1 & \frac{n-i}{2} & \frac{n-i}{2} \\ 
\frac{i-1}{2} & \frac{i+1}{2} & \frac{i-1}{2}-1 & \frac{i-1}{2} \\ 
\frac{n-i}{2} & \frac{n-i}{2}+1 & \frac{n-i}{2}-1 & \frac{n-i}{2}%
\end{array}%
\right)  \label{f6}
\end{equation}

Even $n$, odd $i$: 
\begin{equation}
\left( 
\begin{array}{cccc}
\frac{i+1}{2} & \frac{i+1}{2} & \frac{i-1}{2} & \frac{i-1}{2} \\ 
\frac{n-i+1}{2} & \frac{n-i+1}{2}+1 & \frac{n-i-1}{2} & \frac{n-i+1}{2} \\ 
\frac{i-1}{2} & \frac{i+1}{2} & \frac{i-1}{2}-1 & \frac{i-1}{2} \\ 
\frac{n-i+1}{2} & \frac{n-i+1}{2} & \frac{n-i-1}{2} & \frac{n-i-1}{2}%
\end{array}%
\right)  \label{f7}
\end{equation}

Odd $n$, even $i$: 
\begin{equation}
\left( 
\begin{array}{cccc}
\frac{i}{2} & \frac{i}{2}+1 & \frac{i}{2}-1 & \frac{i}{2} \\ 
\frac{n-i+1}{2} & \frac{n-i+1}{2}+1 & \frac{n-i-1}{2} & \frac{n-i+1}{2} \\ 
\frac{i}{2} & \frac{i}{2} & \frac{i}{2}-1 & \frac{i}{2}-1 \\ 
\frac{n-i+1}{2} & \frac{n-i+1}{2} & \frac{n-i-1}{2} & \frac{n-i-1}{2}%
\end{array}%
\right)  \label{f8}
\end{equation}%
The matrices (\ref{f5} -- \ref{f8}) contain only the arguments of the
initial matrix's elements, for brevity.

By reordering the elements in the rows of the matrices (\ref{f5} -- \ref{f8}%
) we obtain the following new four matrices:

Even $n$, even $i$: 
\begin{equation}
\left( 
\begin{array}{cccc}
\frac{i}{2}-1 & \frac{i}{2} & \frac{i}{2} & \frac{i}{2}+1 \\ 
\frac{n-i}{2}+1 & \frac{n-i}{2} & \frac{n-i}{2} & \frac{n-i}{2}+1 \\ 
\frac{i}{2} & \frac{i}{2}-1 & \frac{i}{2}-1 & \frac{i}{2} \\ 
\frac{n-i}{2} & \frac{n-i}{2}-1 & \frac{n-i}{2}+1 & \frac{n-i}{2}%
\end{array}%
\right)  \label{f9}
\end{equation}

Odd $n$, odd $i$: 
\begin{equation}
\left( 
\begin{array}{cccc}
\frac{i-1}{2} & \frac{i+1}{2} & \frac{i+1}{2} & \frac{i-1}{2} \\ 
\frac{n-i}{2}+1 & \frac{n-i}{2} & \frac{n-i}{2} & \frac{n-i}{2}+1 \\ 
\frac{i-1}{2}-1 & \frac{i-1}{2} & \frac{i-1}{2} & \frac{i+1}{2} \\ 
\frac{n-i}{2} & \frac{n-i}{2}-1 & \frac{n-i}{2}+1 & \frac{n-i}{2}%
\end{array}%
\right)  \label{f10}
\end{equation}

Even $n$, odd $i$: 
\begin{equation}
\left( 
\begin{array}{cccc}
\frac{i-1}{2} & \frac{i+1}{2} & \frac{i+1}{2} & \frac{i-1}{2} \\ 
\frac{n-i+1}{2} & \frac{n-i-1}{2} & \frac{n-i+1}{2}+1 & \frac{n-i+1}{2} \\ 
\frac{i-1}{2}-1 & \frac{i-1}{2} & \frac{i-1}{2} & \frac{i+1}{2} \\ 
\frac{n-i-1}{2} & \frac{n-i+1}{2} & \frac{n-i+1}{2} & \frac{n-i-1}{2}%
\end{array}%
\right)  \label{f11}
\end{equation}

Odd $n$, even $i$: 
\begin{equation}
\left( 
\begin{array}{cccc}
\frac{i}{2}-1 & \frac{i}{2} & \frac{i}{2} & \frac{i}{2}+1 \\ 
\frac{n-i+1}{2} & \frac{n-i-1}{2} & \frac{n-i+1}{2}+1 & \frac{n-i+1}{2} \\ 
\frac{i}{2} & \frac{i}{2}-1 & \frac{i}{2}-1 & \frac{i}{2} \\ 
\frac{n-i-1}{2} & \frac{n-i+1}{2} & \frac{n-i+1}{2} & \frac{n-i-1}{2}%
\end{array}%
\right)  \label{f12}
\end{equation}%
The matrices (\ref{f9} -- \ref{f12}) correspond to the matrices (\ref{f5} -- %
\ref{f8}), respectively. After transforming and reordering the elements in
the columns of the matrices (\ref{f9} -- \ref{f12}), we get:

Even $n$, even $i$: 
\begin{equation}
\left( 
\begin{array}{cccc}
\frac{i}{2} & \frac{i}{2} & \frac{i}{2} & \left( \frac{i}{2}+1\right) \\ 
\frac{n}{2}-\frac{i}{2}+1 & \left( \frac{n}{2}-1\right) -\frac{i}{2}+1 & 
\frac{n}{2}-\frac{i}{2}+1 & \left( \frac{n}{2}+1\right) -\left( \frac{i}{2}%
+1\right) +1 \\ 
\frac{i}{2}-1 & \frac{i}{2}-1 & \frac{i}{2}-1 & \left( \frac{i}{2}+1\right)
-1 \\ 
\frac{n}{2}-\frac{i}{2} & \left( \frac{n}{2}-1\right) -\frac{i}{2} & \frac{n%
}{2}-\frac{i}{2} & \left( \frac{n}{2}+1\right) -\left( \frac{i}{2}+1\right)%
\end{array}%
\right) \qquad \qquad  \label{f15}
\end{equation}

Odd $n$, odd $i$: 
\begin{equation}
\left( 
\begin{array}{cccc}
\frac{i-1}{2} & \frac{i+1}{2} & \frac{i+1}{2} & \frac{i+1}{2} \\ 
\frac{n-1}{2}-\frac{i-1}{2}+1 & \frac{n-1}{2}-\frac{i+1}{2}+1 & \frac{n+1}{2}%
-\frac{i+1}{2}+1 & \frac{n+1}{2}-\frac{i+1}{2}+1 \\ 
\frac{i-1}{2}-1 & \frac{i+1}{2}-1 & \frac{i+1}{2}-1 & \frac{i+1}{2}-1 \\ 
\frac{n-1}{2}-\frac{i-1}{2} & \frac{n-1}{2}-\frac{i+1}{2} & \frac{n+1}{2}-%
\frac{i+1}{2} & \frac{n+1}{2}-\frac{i+1}{2}%
\end{array}%
\right)  \label{f16}
\end{equation}

Even $n$, odd $i$: 
\begin{equation}
\left( 
\begin{array}{cccc}
\frac{i-1}{2} & \frac{i+1}{2} & \frac{i+1}{2} & \frac{i+1}{2} \\ 
\left( \frac{n}{2}-1\right) -\frac{i-1}{2}+1 & \frac{n}{2}-\frac{i+1}{2}+1 & 
\left( \frac{n}{2}+1\right) -\frac{i+1}{2}+1 & \frac{n}{2}-\frac{i+1}{2}+1
\\ 
\frac{i-1}{2}-1 & \frac{i+1}{2}-1 & \frac{i+1}{2}-1 & \frac{i+1}{2}-1 \\ 
\left( \frac{n}{2}-1\right) -\frac{i-1}{2} & \frac{n}{2}-\frac{i+1}{2} & 
\left( \frac{n}{2}+1\right) -\frac{i+1}{2} & \frac{n}{2}-\frac{i+1}{2}%
\end{array}%
\right) \qquad \qquad  \label{f17}
\end{equation}

Odd $n$, even $i$: 
\begin{equation}
\left( 
\begin{array}{cccc}
\frac{i}{2} & \frac{i}{2} & \frac{i}{2} & \left( \frac{i}{2}+1\right) \\ 
\frac{n-1}{2}-\frac{i}{2}+1 & \frac{n-1}{2}-\frac{i}{2}+1 & \frac{n+1}{2}-%
\frac{i}{2}+1 & \frac{n+1}{2}-\left( \frac{i}{2}+1\right) +1 \\ 
\frac{i}{2}-1 & \frac{i}{2}-1 & \frac{i}{2}-1 & \left( \frac{i}{2}+1\right)
-1 \\ 
\frac{n-1}{2}-\frac{i}{2} & \frac{n-1}{2}-\frac{i}{2} & \frac{n+1}{2}-\frac{i%
}{2} & \frac{n+1}{2}-\left( \frac{i}{2}+1\right)%
\end{array}%
\right)  \label{f18}
\end{equation}%
The matrices (\ref{f15} -- \ref{f18}) correspond to the matrices (\ref{f9}
-- \ref{f12}), respectively.

The columns of the matrix (\ref{f15}) (from left to right) are exactly the $%
\varepsilon $-partitions of $2\left( \frac{n}{2}\right) $, $2\left( \frac{n}{%
2}-1\right) $, $2\left( \frac{n}{2}\right) $, and $2\left( \frac{n}{2}%
+1\right) $, respectively. In the same way, the columns of the matrix (\ref%
{f16}) (from left to right) are the $\varepsilon $-partitions of $2\left( 
\frac{n-1}{2}\right) $, $2\left( \frac{n-1}{2}\right) $, $2\left( \frac{n+1}{%
2}\right) $, and $2\left( \frac{n+1}{2}\right) $, respectively; the columns
of the matrix (\ref{f17}) (from left to right) are the $\varepsilon $%
-partitions of $2\left( \frac{n}{2}-1\right) $, $2\left( \frac{n}{2}\right) $%
, $2\left( \frac{n}{2}+1\right) $, and $2\left( \frac{n}{2}\right) $,
respectively; and the columns of the matrix (\ref{f18}) (from left to right)
are the $\varepsilon $-partitions of $2\left( \frac{n-1}{2}\right) $, $%
2\left( \frac{n-1}{2}\right) $, $2\left( \frac{n+1}{2}\right) $, and $%
2\left( \frac{n+1}{2}\right) $, respectively. On the other hand, $%
\left\lceil \frac{n}{2}\right\rceil $, $\left\lfloor \frac{n}{2}%
\right\rfloor +1$, $\left\lceil \frac{n}{2}\right\rceil -1$, and $%
\left\lfloor \frac{n}{2}\right\rfloor $ are equal just to $\frac{n}{2}$, $%
\frac{n}{2}+1$, $\frac{n}{2}-1$, and $\frac{n}{2}$, respectively, for even $%
n $ and to $\frac{n+1}{2}$, $\frac{n+1}{2}$, $\frac{n-1}{2}$, and $\frac{n-1%
}{2}$, respectively, for odd $n$. Thus, columns of each of the matrices (\ref%
{f9} -- \ref{f12}) are the $\varepsilon $-partitions of $2\left\lceil \frac{n%
}{2}\right\rceil $, $2\left( \left\lfloor \frac{n}{2}\right\rfloor +1\right) 
$, $2\left( \left\lceil \frac{n}{2}\right\rceil -1\right) $, and $%
2\left\lfloor \frac{n}{2}\right\rfloor $. Therefore, the necessary groups of
the elements are found in the matrix (\ref{f4}).

However, (\ref{f1}) is not proven yet. As noted above, equations (\ref{f19}
-- \ref{f22}) are correct only for $i\geq 4,$ $n-i\geq 3$. For example, if $%
i=3$, then $i-1=2$ and equation (\ref{f21}) does not hold. For $i=2$ both (%
\ref{f19}) and (\ref{f21}) are not correct. The same can be said about (\ref%
{f20}) and (\ref{f22}). In this case, some of the equations should be
replaced by corresponding inequalities. Indeed, 
\begin{equation}
T\left( \left\lceil \frac{2}{2}\right\rceil \right) +T\left( \left\lfloor 
\frac{2}{2}\right\rfloor +1\right) +T\left( \left\lceil \frac{2}{2}%
\right\rceil -1\right) +T\left( \left\lfloor \frac{2}{2}\right\rfloor
\right) +1=T(2)+1  \label{f23}
\end{equation}%
and 
\begin{equation}
T\left( \left\lceil \frac{1}{2}\right\rceil \right) +T\left( \left\lfloor 
\frac{1}{2}\right\rfloor +1\right) +T\left( \left\lceil \frac{1}{2}%
\right\rceil -1\right) +T\left( \left\lfloor \frac{1}{2}\right\rfloor
\right) +1=T(1)+1  \label{f24}
\end{equation}%
and, therefore, 
\begin{equation}
T(2)<T\left( \left\lceil \frac{2}{2}\right\rceil \right) +T\left(
\left\lfloor \frac{2}{2}\right\rfloor +1\right) +T\left( \left\lceil \frac{2%
}{2}\right\rceil -1\right) +T\left( \left\lfloor \frac{2}{2}\right\rfloor
\right) +1  \label{f25}
\end{equation}%
and 
\begin{equation}
T(1)<T\left( \left\lceil \frac{1}{2}\right\rceil \right) +T\left(
\left\lfloor \frac{1}{2}\right\rfloor +1\right) +T\left( \left\lceil \frac{1%
}{2}\right\rceil -1\right) +T\left( \left\lfloor \frac{1}{2}\right\rfloor
\right) +1.  \label{f26}
\end{equation}%
These relations follow, also, from (\ref{f14}). We consider cases when $%
n\geq 5$ and $2\leq i\leq n-1$. Therefore, $0$, $1$, or $2$ equations from (%
\ref{f19} -- \ref{f22}) should be replaced by inequalities (specifically, $2$
equations for $n=5$; $1$ or $2$ ones for $n=6$; and $0$, or $1$, or $2$ ones
for $n\geq 7$).

Moreover, the opposite assertion for (\ref{f13}) should be proven, i.e., we
have to show that 
\begin{equation}
T(n)<T(i)+T(n-i+1)+T(i-1)+T(n-i)+1  \label{f3}
\end{equation}%
for $n\geq 5$ if $i$ is not equal to $\frac{n+1}{2}$ for odd $n$ and to $%
\frac{n}{2}$ or $\frac{n}{2}+1$ for even $n$ (we will denote these values by 
$n_{/2}$, for brevity). Therefore, it should be shown that the matrix (\ref%
{f4}) contains quartets which provide local strict inequalities for $T\left(
\left\lceil \frac{n}{2}\right\rceil \right) $, $T\left( \left\lfloor \frac{n%
}{2}\right\rfloor +1\right) $, $T\left( \left\lceil \frac{n}{2}\right\rceil
-1\right) $, or $T\left( \left\lfloor \frac{n}{2}\right\rfloor \right) $.
Sums of elements in these quartets added by $1$ should exceed $T\left(
\left\lceil \frac{n}{2}\right\rceil \right) $, or $T\left( \left\lfloor 
\frac{n}{2}\right\rfloor +1\right) $, or $T\left( \left\lceil \frac{n}{2}%
\right\rceil -1\right) $, or $T\left( \left\lfloor \frac{n}{2}\right\rfloor
\right) $, respectively. Corresponding differences have to provide the
correctness of the strict inequality (\ref{f3}) for $i\neq n_{/2}$, even
when relations (\ref{f23} -- \ref{f26}) appear.

The correctness of (\ref{f3}) for $i\neq n_{/2}$ when $n$ is equal to $3$ or 
$4$ has already been shown above. Now we should ascertain how the value of $%
i $ influences the elements of each of the revealed quartets. Suppose that a
column of one of the matrices (\ref{f15} -- \ref{f18}) is an $\varepsilon $%
-partitions of $2k$. Then, the elements of the column are distributed around 
$j$ $(2\leq j\leq k-1)$ in the same way as $i$, $n-i+1$, $i-1$, and $n-i$
are distributed around $i$ $(2\leq i\leq n-1)$. Since $k$ is less than $n$,
a relation like (\ref{f3}) for the column ($n$ and $i$ are replaced by $k$
and $j$, respectively) holds if $j\neq k_{/2}$. Hence, in order for (\ref{f3}%
) to be correct for $i\neq n_{/2}$ there should be at least one column in
each of the matrices (\ref{f15} -- \ref{f18}) in which $j\neq k_{/2}$ when $%
i\neq n_{/2}$.

Therefore, each column in each of these matrices has to be examined (every
column is characterized by its own pair of $k$ and $j$). Besides, every case
should be considered twice: for even $k$ and for odd $k$. For each possible
combination, we suppose that $j=k_{/2}$ and check whether $i=n_{/2}$ in this
case. If $i\neq n_{/2}$, then the corresponding column is recorded as 
\textit{unlucky} (the inequality of $i$ and $n_{/2}$ does not provide the
inequality of $j$ and $k_{/2}$ in this column) and the value of $i$ is
saved. Here are all the tests:

\begin{enumerate}
\item \label{11}Matrix (\ref{f15}) -- even $n$, even $i$

\ref{11}.1. Even $\frac{n}{2}$

\quad Column 1: $k=\frac{n}{2}$ (even)$,$ $j=\frac{i}{2}$

\qquad a) $\frac{i}{2}=\frac{\frac{n}{2}}{2}\Leftrightarrow i=\frac{n}{2}%
=n_{/2}$

\qquad b) $\frac{i}{2}=\frac{\frac{n}{2}}{2}+1\Leftrightarrow i=\frac{n}{2}%
+2\neq n_{/2}$

\quad Column 2: $k=\frac{n}{2}-1$ (odd)$,$ $j=\frac{i}{2}$

$\qquad \frac{i}{2}=\frac{\frac{n}{2}-1+1}{2}\Leftrightarrow i=\frac{n}{2}%
=n_{/2}$

\quad Column 3: $k=\frac{n}{2}$ (even)$,$ $j=\frac{i}{2}$ -- as Column 1

\quad Column 4: $k=\frac{n}{2}+1$ (odd)$,$ $j=\frac{i}{2}+1$

$\qquad \frac{i}{2}+1=\frac{\frac{n}{2}+1+1}{2}\Leftrightarrow i=\frac{n}{2}%
=n_{/2}$

\quad Result: two unlucky columns when $i=\frac{n}{2}+2.$

\ref{11}.2. Odd $\frac{n}{2}$

\quad Column 1: $k=\frac{n}{2}$ (odd)$,$ $j=\frac{i}{2}$

\qquad $\frac{i}{2}=\frac{\frac{n}{2}+1}{2}\Leftrightarrow i=\frac{n}{2}%
+1=n_{/2}$

\quad Column 2: $k=\frac{n}{2}-1$ (even)$,$ $j=\frac{i}{2}$

$\qquad $a) $\frac{i}{2}=\frac{\frac{n}{2}-1}{2}\Leftrightarrow i=\frac{n}{2}%
-1\neq n_{/2}$

$\qquad $b) $\frac{i}{2}=\frac{\frac{n}{2}-1}{2}+1\Leftrightarrow i=\frac{n}{%
2}+1=n_{/2}$

\quad Column 3: $k=\frac{n}{2}$ (odd)$,$ $j=\frac{i}{2}$ -- as Column 1

\quad Column 4: $k=\frac{n}{2}+1$ (even)$,$ $j=\frac{i}{2}+1$

$\qquad $a) $\frac{i}{2}+1=\frac{\frac{n}{2}+1}{2}\Leftrightarrow i=\frac{n}{%
2}-1\neq n_{/2}$

$\qquad $b) $\frac{i}{2}+1=\frac{\frac{n}{2}+1}{2}+1\Leftrightarrow i=\frac{n%
}{2}+1=n_{/2}$

\quad Result: two unlucky columns when $i=\frac{n}{2}-1.$

\item \label{22}Matrix (\ref{f16}) -- odd $n$, odd $i$

\ref{22}.1. Even $\frac{n-1}{2}$

\quad Column 1: $k=\frac{n-1}{2}$ (even)$,$ $j=\frac{i-1}{2}$

\qquad a) $\frac{i-1}{2}=\frac{\frac{n-1}{2}}{2}\Leftrightarrow i=\frac{n+1}{%
2}=n_{/2}$

\qquad b) $\frac{i-1}{2}=\frac{\frac{n-1}{2}}{2}+1\Leftrightarrow i=\frac{n+5%
}{2}\neq n_{/2}$

\quad Column 2: $k=\frac{n-1}{2}$ (even)$,$ $j=\frac{i+1}{2}$

$\qquad $a) $\frac{i+1}{2}=\frac{\frac{n-1}{2}}{2}\Leftrightarrow i=\frac{n-3%
}{2}\neq n_{/2}$

$\qquad $b) $\frac{i+1}{2}=\frac{\frac{n-1}{2}}{2}+1\Leftrightarrow i=\frac{%
n+1}{2}=n_{/2}$

\quad Column 3: $k=\frac{n+1}{2}$ (odd)$,$ $j=\frac{i+1}{2}$

\qquad $\frac{i+1}{2}=\frac{\frac{n+1}{2}+1}{2}\Leftrightarrow i=\frac{n+1}{2%
}=n_{/2}$

\quad Column 4: $k=\frac{n+1}{2}$ (odd)$,$ $j=\frac{i+1}{2}$ -- as Column 3

\quad Result: one unlucky column when $i=\frac{n+5}{2}$ and one unlucky
column when $i=\frac{n-3}{2}.$

\ref{22}.2. Odd $\frac{n-1}{2}$

\quad Column 1: $k=\frac{n-1}{2}$ (odd)$,$ $j=\frac{i-1}{2}$

\qquad $\frac{i-1}{2}=\frac{\frac{n-1}{2}+1}{2}\Leftrightarrow i=\frac{n+3}{2%
}\neq n_{/2}$

\quad Column 2: $k=\frac{n-1}{2}$ (odd)$,$ $j=\frac{i+1}{2}$

$\qquad \frac{i+1}{2}=\frac{\frac{n-1}{2}+1}{2}\Leftrightarrow i=\frac{n-1}{2%
}\neq n_{/2}$

\quad Column 3: $k=\frac{n+1}{2}$ (even)$,$ $j=\frac{i+1}{2}$

\qquad a) $\frac{i+1}{2}=\frac{\frac{n+1}{2}}{2}\Leftrightarrow i=\frac{n-1}{%
2}\neq n_{/2}$

\qquad b) $\frac{i+1}{2}=\frac{\frac{n+1}{2}}{2}+1\Leftrightarrow i=\frac{n+3%
}{2}\neq n_{/2}$

\quad Column 4: $k=\frac{n+1}{2}$ (even)$,$ $j=\frac{i+1}{2}$ -- as Column 3

\quad Result: three unlucky columns when $i=\frac{n+3}{2}$ and three unlucky
columns when $i=\frac{n-1}{2}.$

\item \label{33}Matrix (\ref{f17}) -- even $n$, odd $i$

\ref{33}.1. Even $\frac{n}{2}$

\quad Column 1: $k=\frac{n}{2}-1$ (odd)$,$ $j=\frac{i-1}{2}$

\qquad $\frac{i-1}{2}=\frac{\frac{n}{2}-1+1}{2}\Leftrightarrow i=\frac{n}{2}%
+1=n_{/2}$

\quad Column 2: $k=\frac{n}{2}$ (even)$,$ $j=\frac{i+1}{2}$

$\qquad $a) $\frac{i+1}{2}=\frac{\frac{n}{2}}{2}\Leftrightarrow i=\frac{n}{2}%
-1\neq n_{/2}$

$\qquad $b) $\frac{i+1}{2}=\frac{\frac{n}{2}}{2}+1\Leftrightarrow i=\frac{n}{%
2}+1=n_{/2}$

\quad Column 3: $k=\frac{n}{2}+1$ (odd)$,$ $j=\frac{i+1}{2}$

\qquad $\frac{i+1}{2}=\frac{\frac{n}{2}+1+1}{2}\Leftrightarrow i=\frac{n}{2}%
+1=n_{/2}$

\quad Column 4: $k=\frac{n}{2}$ (even)$,$ $j=\frac{i+1}{2}$ -- as Column 2

\quad Result: two unlucky columns when $i=\frac{n}{2}-1.$

\ref{33}.2. Odd $\frac{n}{2}$

\quad Column 1: $k=\frac{n}{2}-1$ (even)$,$ $j=\frac{i-1}{2}$

\qquad a) $\frac{i-1}{2}=\frac{\frac{n}{2}-1}{2}\Leftrightarrow i=\frac{n}{2}%
=n_{/2}$

\qquad b) $\frac{i-1}{2}=\frac{\frac{n}{2}-1}{2}+1\Leftrightarrow i=\frac{n}{%
2}+2\neq n_{/2}$

\quad Column 2: $k=\frac{n}{2}$ (odd)$,$ $j=\frac{i+1}{2}$

$\qquad \frac{i+1}{2}=\frac{\frac{n}{2}+1}{2}\Leftrightarrow i=\frac{n}{2}%
=n_{/2}$

\quad Column 3: $k=\frac{n}{2}+1$ (even)$,$ $j=\frac{i+1}{2}$

\qquad a) $\frac{i+1}{2}=\frac{\frac{n}{2}+1}{2}\Leftrightarrow i=\frac{n}{2}%
=n_{/2}$

\qquad b) $\frac{i+1}{2}=\frac{\frac{n}{2}+1}{2}+1\Leftrightarrow i=\frac{n}{%
2}+2\neq n_{/2}$

\quad Column 4: $k=\frac{n}{2}$ (odd)$,$ $j=\frac{i+1}{2}$ -- as Column 2

\quad Result: two unlucky columns when $i=\frac{n}{2}+2.$

\item \label{44}Matrix (\ref{f18}) -- odd $n$, even $i$

\ref{44}.1. Even $\frac{n-1}{2}$

\quad Column 1: $k=\frac{n-1}{2}$ (even)$,$ $j=\frac{i}{2}$

\qquad a) $\frac{i}{2}=\frac{\frac{n-1}{2}}{2}\Leftrightarrow i=\frac{n-1}{2}%
\neq n_{/2}$

\qquad b) $\frac{i}{2}=\frac{\frac{n-1}{2}}{2}+1\Leftrightarrow i=\frac{n-1}{%
2}+2=\frac{n+3}{2}\neq n_{/2}$

\quad Column 2: $k=\frac{n-1}{2}$ (even)$,$ $j=\frac{i}{2}$ - as Column 1

\quad Column 3: $k=\frac{n+1}{2}$ (odd)$,$ $j=\frac{i}{2}$

\qquad $\frac{i}{2}=\frac{\frac{n+1}{2}+1}{2}\Leftrightarrow i=\frac{n+1}{2}%
+1=\frac{n+3}{2}\neq n_{/2}$

\quad Column 4: $k=\frac{n+1}{2}$ (odd)$,$ $j=\frac{i}{2}+1$

\qquad $\frac{i}{2}+1=\frac{\frac{n+1}{2}+1}{2}\Leftrightarrow i=\frac{n+1}{2%
}-1=\frac{n-1}{2}\neq n_{/2}$

\quad Result: three unlucky columns when $i=\frac{n-1}{2}$ and three unlucky
columns when $i=\frac{n+3}{2}.$

\ref{44}.2. Odd $\frac{n-1}{2}$

\quad Column 1: $k=\frac{n-1}{2}$ (odd)$,$ $j=\frac{i}{2}$

\qquad $\frac{i}{2}=\frac{\frac{n-1}{2}+1}{2}\Leftrightarrow i=\frac{n+1}{2}%
=n_{/2}$

\quad Column 2: $k=\frac{n-1}{2}$ (odd)$,$ $j=\frac{i}{2}$ - as Column 1

\quad Column 3: $k=\frac{n+1}{2}$ (even)$,$ $j=\frac{i}{2}$

\qquad a) $\frac{i}{2}=\frac{\frac{n+1}{2}}{2}\Leftrightarrow i=\frac{n+1}{2}%
=n_{/2}$

\qquad b) $\frac{i}{2}=\frac{\frac{n+1}{2}}{2}+1\Leftrightarrow i=\frac{n+1}{%
2}+2=\frac{n+5}{2}\neq n_{/2}$

\quad Column 4: $k=\frac{n+1}{2}$ (even)$,$ $j=\frac{i}{2}+1$

\qquad a) $\frac{i}{2}+1=\frac{\frac{n+1}{2}}{2}\Leftrightarrow i=\frac{n+1}{%
2}-2=\frac{n-3}{2}\neq n_{/2}$

\qquad b) $\frac{i}{2}+1=\frac{\frac{n+1}{2}}{2}+1\Leftrightarrow i=\frac{n+1%
}{2}=n_{/2}$

\quad Result: one unlucky column when $i=\frac{n+5}{2}$ and one unlucky
column when $i=\frac{n-3}{2}.$
\end{enumerate}

Hence, the tests show that for any $n$ and $i$ there exists at least one
column in the matrix (\ref{f4}) that provides a local strict inequality.
Moreover, unlucky columns appear only for a limited set of values of $i$.
All these values are close to $n_{/2}$. For this reason, in the general
case, for large $n$, these values do not coincide with $2$, $3$, $n-2$, and $%
n-1$, i.e., with the \textit{extreme values}, for which some of equations (%
\ref{f19} -- \ref{f22}) are replaced by inequalities (\ref{f25}) or (\ref%
{f26}). As follows from (\ref{f23}) and (\ref{f24}) the difference between
the right and the left parts of (\ref{f25}) and (\ref{f26}), respectively,
is equal to one. Hence, the total difference given by (\ref{f25}) and (\ref%
{f26}) with the extreme values of $i$ is equal to two. This difference
decreases the difference between the right and the left parts of (\ref{f3}).
On the other hand, four columns of the matrix (\ref{f4}) provide four local
strict inequalities in this a case. These inequalities give a total
difference with an opposite sign and its absolute value is equal to four at
least. This difference increases the difference between the right and the
left parts of (\ref{f3}). Hence, the strict inequality (\ref{f3}) holds for
extreme values of $i$. On the other hand, if $i$ is close to $n_{/2}$ then,
as noted above, at least one column is not unlucky and provides a local
strict inequality. Hence, (\ref{f3}) holds also in this case. All the more, (%
\ref{f3}) is correct for all other values of $i$.

We consider now the special cases when the extreme values of $i$ are close
to $n_{/2}$. These cases take place for small $n$. Therefore, correctness of
(\ref{f3}) should be checked separately for corresponding combinations of $n$
and $i$. According to (\ref{f2}), the following relations hold: \bigskip 
\begin{eqnarray*}
T(5) &=&T(3)+T(3)+T(2)+T(2)+1 \\
&=&3+3+2+2+1=9 \\
T(6) &=&T(3)+T(4)+T(2)+T(3)+1 \\
&=&3+6+1+3+1=14 \\
T(7) &=&T(4)+T(4)+T(3)+T(3)+1 \\
&=&6+6+3+3+1=19 \\
T(8) &=&T(4)+T(5)+T(3)+T(4)+1 \\
&=&6+9+3+6+1=25.
\end{eqnarray*}%
We begin from the cases when three unlucky columns exist. As shown above
three unlucky columns appear for odd $n$, when $i=\frac{n+3}{2}$ (i.e., odd $%
i$ and odd $\frac{n-1}{2}$) and when $i=\frac{n-1}{2}$ (i.e., even $i$ and
even $\frac{n-1}{2}$). In the considered range of $n$, these values of $i$
coincide with the extreme values when $n=7$, in the first case, and when $%
n=5 $, in the second case. Indeed, for $n=7$ 
\begin{eqnarray*}
i &=&\frac{n+3}{2}=\frac{7+3}{2}=5=n-2 \\
i &=&\frac{n-1}{2}=\frac{7-1}{2}=3
\end{eqnarray*}%
and for $n=5$ 
\begin{eqnarray*}
i &=&\frac{n+3}{2}=\frac{5+3}{2}=4=n-1 \\
i &=&\frac{n-1}{2}=\frac{5-1}{2}=2.
\end{eqnarray*}%
For both $i=n-2$ and $i=3$ the right part of (\ref{f3}), with $n=7$ is 
\begin{equation*}
T(3)+T(5)+T(2)+T(4)+1=3+9+1+6+1=20>T(7)
\end{equation*}
and for both $i=n-1$ and $i=2$ the right part of (\ref{f3}) with $n=5$ is 
\begin{equation*}
T(2)+T(4)+T(1)+T(3)+1=1+6+0+3+1=11>T(5).
\end{equation*}
Now consider cases when two unlucky columns exist. As shown above two
unlucky columns appear for even $n$, when $i=\frac{n}{2}+2$ (even $i$, even $%
\frac{n}{2}$ or odd $i$, odd $\frac{n}{2}$) and when $i=\frac{n}{2}-1$ (even 
$i$, odd $\frac{n}{2}$ or odd $i$, even $\frac{n}{2}$). In the considered
range of $n$, these values of $i$ coincide with the extreme values when $n=8$%
, and when $n=6$. Indeed, for $n=8$ 
\begin{eqnarray*}
i &=&\frac{n}{2}+2=\frac{8}{2}+2=6=n-2 \\
i &=&\frac{n}{2}-1=\frac{8}{2}-1=3
\end{eqnarray*}%
and for $n=6$%
\begin{eqnarray*}
i &=&\frac{n}{2}+2=\frac{6}{2}+2=5=n-1 \\
i &=&\frac{n}{2}-1=\frac{6}{2}-1=2.
\end{eqnarray*}%
For both $i=n-2$ and $i=3$ the right part of (\ref{f3}), with $n=8$ is 
\begin{equation*}
T(3)+T(6)+T(2)+T(5)+1=3+14+1+9+1=28>T(8)
\end{equation*}
and for both $i=n-1$ and $i=2$ the right part of (\ref{f3}) with $n=6$ is 
\begin{equation*}
T(2)+T(5)+T(1)+T(4)+1=1+9+0+6+1=17>T(6).
\end{equation*}
We need not consider separately cases with one unlucky column, since then
three columns of the matrix (\ref{f4}) provide three local strict
inequalities. These inequalities give a total difference that increases the
difference between the right and the left parts of (\ref{f3}) by three at
least. As noted above, inequalities (\ref{f25}) or (\ref{f26}) can decrease
the difference by not more than two. Hence, (\ref{f3}) holds in this case,
and the proof of the lemma is complete. 
\endproof%

\begin{remark}
It is of interest to trace why the strict inequality (\ref{f3}) holds in
some special cases when the extreme values of $i$ are close to $n_{/2}$,
specifically, for three unlucky columns. For example, consider $n=7$, $i=3$.
The matrix (\ref{f16}) turns into the right matrix (\ref{f27}) (see below).
For $n=7$, $i=n_{/2}=4$ the matrix (\ref{f18}) turns into the left matrix (%
\ref{f27}). 
\begin{equation}
\left( 
\begin{array}{cccc}
2 & 2 & 2 & 3 \\ 
2 & 2 & 3 & 2 \\ 
1 & 1 & 1 & 2 \\ 
1 & 1 & 2 & 1%
\end{array}%
\right) \qquad \qquad \left( 
\begin{array}{cccc}
1 & 2 & 2 & 2 \\ 
3 & 2 & 3 & 3 \\ 
0 & 1 & 1 & 1 \\ 
2 & 1 & 2 & 2%
\end{array}%
\right)  \label{f27}
\end{equation}%
As expected, three columns of the left matrix coincide with three columns of
the right matrix (up to the order of the elements in the columns). Only the
left columns of the left and the right matrices differ. As noted above,
equation (\ref{f21}) is replaced by an inequality for $i=3$. Hence, the
difference between the right and the left parts of (\ref{f3}) decreases by
one. Corresponding computations along the left columns of the left and the
right matrices (\ref{f27}) give the following results, respectively: 
\begin{eqnarray*}
T(2)+T(2)+T(1)+T(1)+1 &=&1+1+0+0+1=3\text{ } \\
T(1)+T(3)+T(0)+T(2)+1 &=&0+3+0+1+1=5.
\end{eqnarray*}%
The difference between the results is equal to two, i.e., the difference
between the right and the left parts of (\ref{f3}) increases by two.
Therefore, the total difference between the right and the left parts of (\ref%
{f3}) is equal to one in this case, exactly as shown in the proof of Lemma %
\ref{lem_n/2_a}. Other special cases are analyzed similarly.
\end{remark}

\section{The Second Decomposition Lemma}

We define the following recursive function $P(n)$: 
\begin{eqnarray}
P(0) &=&0  \notag \\
P(1) &=&0  \notag \\
P(2) &=&0  \notag \\
P(n) &=&P\left( \left\lceil \frac{n}{2}\right\rceil \right) +P\left(
\left\lfloor \frac{n}{2}\right\rfloor +1\right) +P\left( \left\lceil \frac{n%
}{2}\right\rceil -1\right) +P\left( \left\lfloor \frac{n}{2}\right\rfloor
\right) +1\qquad \qquad  \label{f28} \\
(n &\geq &3).  \notag
\end{eqnarray}

\begin{lemma}
\label{lem_n/2_P_a}For $n\geq 1$, $1\leq i\leq n$%
\begin{equation}
P(n)\leq P(i)+P(n-i+1)+P(i-1)+P(n-i)+1.  \label{f29}
\end{equation}
\end{lemma}

\proof%
The structure of the general equation for $P(n)$ in (\ref{f28}) is the same
as the structure of the general equation for $T(n)$ in (\ref{f2}). Thereby,
the general construction of the proof is the same as in Lemma \ref{lem_n/2_a}%
. The substitution of $n=2$ and $n=1$ gives the same equations and
inequalities for $P(n)$ as in (\ref{f23} -- \ref{f26}) for $T(n)$. Hence,
the special cases only, when extreme values of $i$ are close to $n_{/2}$,
need to be checked separately. The statement that is proven is weaker than
the statement in Lemma \ref{lem_n/2_a}. We need not prove the inequality as (%
\ref{f3}) for $i\neq n_{/2}$. For this reason, the cases with two unlucky
columns in the matrix (\ref{f4}) need not be considered. Indeed, in these
cases the rest of the columns provide two local strict inequalities. These
inequalities give a total difference that increases the difference between
the right and the left parts of (\ref{f29}) by two at least. On the other
hand, inequalities as (\ref{f25}) and (\ref{f26}) can decrease the
difference by not more than two. Hence, (\ref{f29}) will hold in these
cases. Therefore, the cases with three unlucky columns only have to be
considered. As shown in the proof of Lemma \ref{lem_n/2_a}, there are two
such cases. In the first case, $n=7$ and $i$ is equal to $3$ or $5$. In the
second case, $n=5$ and $i$ is equal to $2$ or $4$. According to (\ref{f28}), 
\begin{eqnarray*}
P(3) &=&P(2)+P(2)+P(1)+P(1)+1 \\
&=&0+0+0+0+1=1 \\
P(4) &=&P(2)+P(3)+P(1)+P(2)+1 \\
&=&0+1+0+0+1=2 \\
P(5) &=&P(3)+P(3)+P(2)+P(2)+1 \\
&=&1+1+0+0+1=3 \\
P(7) &=&P(4)+P(4)+P(3)+P(3)+1 \\
&=&2+2+1+1+1=7.
\end{eqnarray*}%
In the first case, the right part of (\ref{f29}) turns into 
\begin{equation*}
P(3)+P(5)+P(2)+P(4)+1=1+3+0+2+1=7=P(7).
\end{equation*}
In the second case, the right part of (\ref{f29}) turns into 
\begin{equation*}
P(2)+P(4)+P(1)+P(3)+1=0+2+0+1+1=4>P(5).
\end{equation*}
Thus, (\ref{f28}) holds in both special cases and the proof of the lemma is
complete. 
\endproof%
\medskip

Hence, there are cases when not only values $n_{/2}$ of $i$ provide the
equality for (\ref{f29}). We intend to investigate all such cases. We begin
from the special cases, when extreme values of $i$ are close to $n_{/2}$. It
is clear that the number of unlucky columns in the matrix (\ref{f4}) should
be not less than two in these cases. Situations with three unlucky columns
were checked in the proof of Lemma \ref{lem_n/2_P_a}. As shown in the proof
of Lemma \ref{lem_n/2_a}, there are two special cases with two unlucky
columns. In the first case, $n=8$ and $i$ is equal to $3$ or $6$. In the
second case, $n=6$ and $i$ is equal to $2$ or $5$. According to (\ref{f28}), 
\begin{eqnarray*}
P(6) &=&P(3)+P(4)+P(2)+P(3)+1 \\
&=&1+2+0+1+1=5 \\
P(8) &=&P(4)+P(5)+P(3)+P(4)+1 \\
&=&2+3+1+2+1=9.
\end{eqnarray*}%
In the first case, the right part of (\ref{f29}) turns into 
\begin{equation*}
P(3)+P(6)+P(2)+P(5)+1=1+5+0+3+1=10>P(8).
\end{equation*}
In the second case, the right part of (\ref{f29}) turns into 
\begin{equation*}
P(2)+P(5)+P(1)+P(4)+1=0+3+0+2+1=6>P(6).
\end{equation*}
Therefore, $7$ is the only value of $n$ that can provide the equality for (%
\ref{f29}) with $i$ that is not equal to $n_{/2}$ (\textit{special value of }%
$n$ or \textit{special number}). Corresponding values of $i$ are $3$ or $5$,
i.e., $\frac{n-1}{2}$ or $\frac{n+3}{2}$, respectively.

Now, situations when any columns of the matrices (\ref{f15} -- \ref{f18})
are $\varepsilon $-partitions of $2\times 7$ have to be examined. These
columns do not guarantee local strict inequalities (\textit{special columns}%
). The combination of special columns with possible unlucky columns can lead
to equalities in (\ref{f29}) when $i\neq n_{/2}$ for new values of $n$.
Several values of $n$ in the left part of general equation (\ref{f28}) give
the appearance of $7$ as an argument in the right part. The following
corresponding decompositions are possible: 
\begin{eqnarray*}
n &=&12:6,7,5,6 \\
n &=&13:7,7,6,6 \\
n &=&14:7,8,6,7 \\
n &=&15:8,8,7,7 \\
n &=&16:8,9,7,8.
\end{eqnarray*}%
As shown in the proof of Lemma \ref{lem_n/2_a}, matrices for even $n$ can
have not more than two unlucky columns. Hence, matrices for $12$ and $16$
can have not more than two unlucky columns and one special column. That is,
there exists at least one \textit{lucky column }providing a local strict
inequality in these cases and, therefore, $12$ and $16$ cannot be a special
numbers. Matrices for $13$, $14$, and $15$ have two special columns. We
should perform corresponding computations for the values of $i$ which give
more than one unlucky column (they are $\frac{n-1}{2}$ or $\frac{n+3}{2}$
for odd $n$ and $\frac{n}{2}$ or $\frac{n}{2}+1$ for even $n$ --- see the
tests of the columns in the proof of Lemma \ref{lem_n/2_a}) and to compare
the results with $P(13)$, $P(14)$, and $P(15)$, respectively: 
\begin{eqnarray*}
P(13) &=& \\
P(7)+P(7)+P(6)+P(6)+1 &=&7+7+5+5+1=25 \\
P(6)+P(8)+P(5)+P(7)+1 &=&5+9+3+7+1=25=P(13) \\
P(14) &=& \\
P(7)+P(8)+P(6)+P(7)+1 &=&7+9+5+7+1=29 \\
P(6)+P(9)+P(5)+P(8)+1 &=&5+11+3+9+1=29=P(14) \\
P(15) &=& \\
P(8)+P(8)+P(7)+P(7)+1 &=&9+9+7+7+1=33 \\
P(7)+P(9)+P(6)+P(8)+1 &=&7+11+5+9+1=33=P(15).
\end{eqnarray*}%
Hence, $13$, $14$, and $15$ are the special numbers also. In principle, it
can be shown strictly, by the substitution of corresponding values of $i$ in
the tests of the columns in the proof of Lemma \ref{lem_n/2_a}. The simple
computations show that just values of $i$ which give more than one unlucky
column provide such $\varepsilon $-partitions of $2\times 7$ in the
corresponding special columns of the matrices (\ref{f15} -- \ref{f18}) that $%
j$ is equal to $3$ or $5$.

Hence, special numbers multiply. Indeed, special columns which are $%
\varepsilon $-partitions of $2\times 13$, $2\times 14$, and $2\times 15$ are
the base for appearance of new special numbers. We have the following
decompositions: 
\begin{eqnarray*}
n &=&24:12,13,11,12 \\
n &=&25:13,13,12,12 \\
n &=&26:13,14,12,13 \\
n &=&27:14,14,13,13 \\
n &=&28:14,15,13,14 \\
n &=&29:15,15,14,14 \\
n &=&30:15,16,14,15 \\
n &=&31:16,16,15,15 \\
n &=&32:16,17,15,16.
\end{eqnarray*}%
Matrices for $24$ and $32$ have a single special column, and, therefore, $24$
and $32$ cannot be special numbers. Matrices for other numbers have from two
to four special columns. Corresponding computations show that all values of $%
n$ from $25$ to $31$ are the special numbers. Substituting the neighbors of $%
n_{/2}$ for $i$ provides the equality for (\ref{f29}) in these cases.
However, for some values of $n$, not only the nearest neighbors of $n_{/2}$
provide the equality for (\ref{f29}). The range of such values of $i$
increases in the middle of a group of special numbers. For example, for $%
n=27 $, the right part of (\ref{f29}) is equal to the same number (it is $%
109 $) when $i$ equals $14$ (it is $\frac{n+1}{2}$, i.e., $n_{/2}$), $13$ or 
$15$ ($\frac{n-1}{2}$ or $\frac{n+3}{2}$), and $12$ or $16$ ($\frac{n-3}{2}$
or $\frac{n+5}{2}$). For $n=29$, the situation is analogous (the result is $%
125$ in all five cases). For $n=28$, the right part of (\ref{f29}) is equal
to the same number (it is $117$) when $i$ equals $14$ or $15$ ($\frac{n}{2}$
or $\frac{n}{2}+1$, i.e., $n_{/2}$), $13$ or $16$ ($\frac{n}{2}-1$ or $\frac{%
n}{2}+2$), and $12$ or $16$ ($\frac{n}{2}-2$ or $\frac{n}{2}+3$). Notice,
that in the odd case, when $i$ is equal to $\frac{n-3}{2}$ or $\frac{n+5}{2}$%
, and in the even case, when $i$ is equal to $\frac{n}{2}-2$ or $\frac{n}{2}%
+3$, there are no unlucky columns at all. All four columns are special in
these cases and present the $\varepsilon $-partitions which provide
corresponding equalities.

The induction method shows that the spreading around $n_{/2}$ of the values
of $i$ which provide the equality in (\ref{f29}) takes place for the
infinite number of successively increasing groups of successive special
values of $n$. The special values of $n$ are grouped as follows: 
\begin{equation*}
7,13\div 15,25\div 31,49\div 63,97\div 127,193\div 255,\ldots
\end{equation*}%
In the general view, they can be presented in the following way: 
\begin{eqnarray*}
n_{first_{\nu }} &\leq &n_{sp_{\nu }}\leq n_{last_{\nu }}, \\
n_{first_{1}} &=&n_{last_{1}}=7, \\
n_{first_{\nu }} &=&2n_{first_{\nu -1}}-1, \\
n_{last_{\nu }} &=&2n_{last_{\nu -1}}+1.
\end{eqnarray*}%
Here $\nu $ is a number of a group of special numbers; $n_{sp_{\nu }}$ is a
special number of the $\nu $-th group; $n_{first_{\nu }}$ and $n_{last_{\nu
}}$ are the first value and the last value, respectively, in the $\nu $-th
group. For all these values of $n$, not only values $n_{/2}$ of $i$ provide
the equality for (\ref{f29}). Other possible values of $i$ are concentrated
around $n_{/2}$. The range of such values of $i$ increases with the approach
to the middle of a group of special values of $n$. On the borders of a
group, these values of $i$ are only $n_{/2}$ and their nearest neighbors.
The possible range of corresponding values of $i$ increases in a transition
from a given group to the next one.

\end{document}